\documentclass[superscriptaddress,showpacs]{revtex4}
\usepackage{epsfig}
\usepackage{times}
\usepackage{amsmath}
\usepackage{amsfonts}
\usepackage{amssymb}
\usepackage[usenames]{color}
\usepackage{graphicx}
\usepackage{bm}
\newcommand{\beq}{\begin{equation}}
\newcommand{\eeq}{\end{equation}}
\newcommand{\bea}{\begin{eqnarray}}
\newcommand{\eea}{\end{eqnarray}}
\newcommand{\bec}{\begin{center}}
\newcommand{\enc}{\end{center}}
\newcommand{\bfr}{\begin{flushright}}
\newcommand{\efr}{\end{flushright}}

\newcommand{\la}{\langle}
\newcommand{\ra}{\rangle}

\newcommand{\om}{\omega}

\newcommand{\g}{\gamma}
\newcommand{\gam}{\gamma}
\newcommand{\s}{\sigma}

\newcommand{\Om}{\Omega}

\newcommand{\Gam}{\Gamma}
\newcommand{\cH}{{\cal H}}

\begin{document}
\title{
Observation of three-state dressed states in circuit quantum electrodynamics
}
\author{K. Koshino}
\affiliation{College of Liberal Arts and Sciences, Tokyo Medical and Dental
University, Ichikawa, Chiba 272-0827, Japan}
\author{H. Terai}
\affiliation{National Institute of Information and Communications Technology (NICT), 
Kobe, Hyogo 651-2492, Japan}
\author{K. Inomata}
\affiliation{The Institute of Physical and Chemical Research (RIKEN), 
Wako, Saitama 351-0198, Japan}
\author{T. Yamamoto}
\affiliation{The Institute of Physical and Chemical Research (RIKEN), 
Wako, Saitama 351-0198, Japan}
\affiliation{Smart Energy Research Laboratories, NEC Corporation,  
Tsukuba, Ibaraki 305-8501, Japan}
\author{W. Qiu}
\affiliation{National Institute of Information and Communications Technology (NICT), 
Kobe, Hyogo 651-2492, Japan}
\author{Z. Wang}
\affiliation{National Institute of Information and Communications Technology (NICT), 
Kobe, Hyogo 651-2492, Japan}
\author{Y. Nakamura}
\affiliation{The Institute of Physical and Chemical Research (RIKEN), 
Wako, Saitama 351-0198, Japan}
\affiliation{Research Center for Advanced Science and Technology (RCAST), 
The University of Tokyo, Meguro-ku, Tokyo 153-8904, Japan}

\date{\today}
\begin{abstract}
We have investigated the microwave response of a transmon qubit
coupled directly to a transmission line.
In a transmon qubit, owing to its weak anharmonicity,
a single driving field may generate 
dressed states involving more than two bare states. 
We confirmed the formation of three-state dressed states
by observing all of the six associated Rabi sidebands,
which appear as either amplification or attenuation of the probe field.
The experimental results are reproduced with good precision
by a theoretical model incorporating 
the radiative coupling between the qubit and the microwave.
\end{abstract}
\pacs{42.50.Gy, 42.50.Ct, 85.25.Cp}
\maketitle

The optical response of two-level systems is a central research subject 
in the field of atomic, molecular and optical physics. 
When two-level systems are driven by a resonant field,
the ground and excited states are mixed to form dressed states.
For each unique transition between dressed states, 
two Rabi sidebands appear symmetrically 
about the transition frequency
in the fluorescence power spectrum~\cite{Mollow1,Mollow2}.
Driving transitions also populates the excited state.
However, even in the strong-driving limit,
a continuous field cannot induce population inversion,
and therefore effects which rely on inversion, such as lasing,
are generally thought to be forbidden in two-level systems.
However, when an additional field, e.g., a probe field, 
is also applied to such a driven atom, 
the field is amplified when tuned to one of the Rabi sidebands,
whereas it is attenuated when tuned to the other sideband. 
In other words, amplification (lasing) 
can take place without explicit population inversion 
in the bare-state basis~\cite{LWI1,LWI2,LWI3,LWI4,LWI5},
provided there is a mechanism for
population inversion in the dressed-state basis.

Many experiments analogous to those in optics 
are now being performed within the context of
circuit quantum electrodynamics (QED),
in which the optical fields and atoms of conventional QED
are replaced with microwave fields
and superconducting qubits, respectively~\cite{cQED1,cQED2}.
An advantage of using superconducting qubits is
their large transition dipole moments,
which enable strong coupling to the microwave field~\cite{cQED3}.
Exploiting this strong coupling,
lasing experiments become possible by using a single quantum emitter:
Maser operation has been demonstrated
by a single driven qubit-cavity system~\cite{lasing1,lasing2}
and a driven qubit coupled directly to a transmission line~\cite{onchip},
where population inversion takes place in the bare-state basis.
Amplification due to dressed-state inversion
has been reported recently in a qubit-cavity system~\cite{dsa}, 
and lasing without inversion independent of basis 
has also been proposed theoretically~\cite{LWI10}.
Furthermore, the Autler-Townes effect and 
the electromagnetically induced transparency have been investigated
by driving a higher transition of the qubit~\cite{EIT1,EIT2,EIT3,EIT4,EIT5,EIT6},
and the quantum interference induced by 
longitudinal driving has been confirmed~\cite{LZ1,LZ2,LZ3}.

Superconducting qubits are multilevel quantum systems having 
designed energy-level spacings.
In transmon qubits, 
the eigenstates of interest are formed at the bottom of a cosine potential 
originating from the Josephson energy, 
which is much larger than the single-electron charging energy. 
Because of the potential's weak anharmonicity, 
the level spacings are nearly degenerate, although unique.~\cite{tra}. 
In previous works using transmon qubits,
the lowest transition was used as a two-level system
and its strong sensitivity to single microwave photons
was confirmed~\cite{NL1,NL2}.
In this work, we demonstrate a phenomenon in which
the weak anharmonicity of a transmon qubit is essential:
Owing to the close transition frequencies
and the large transition dipole moments,
a single drive wave can excite
not only the lowest transition ($|0\ra\leftrightarrow|1\ra$)
but also higher ones ($|1\ra\leftrightarrow|2\ra$, etc.) simultaneously. 
As a result, the dressed states may involve more than two bare states.
Here we report the observation of three-state dressed states
and the six resulting Rabi sidebands,
which are manifest in our experiment as the
amplification or attenuation of a probe field resonant with those sidebands.
These experimental results are reproduced 
by a simple theoretical model accounting for
the radiative coupling between the qubit and the transmission line. 
The realization of coherent quantum dynamics
in a dressed-state basis 
induced by microwave photons is an important step towards 
constructing scalable quantum networks~\cite{rtswap,OE}.

\begin{figure*}
\bec\includegraphics[width=160mm]{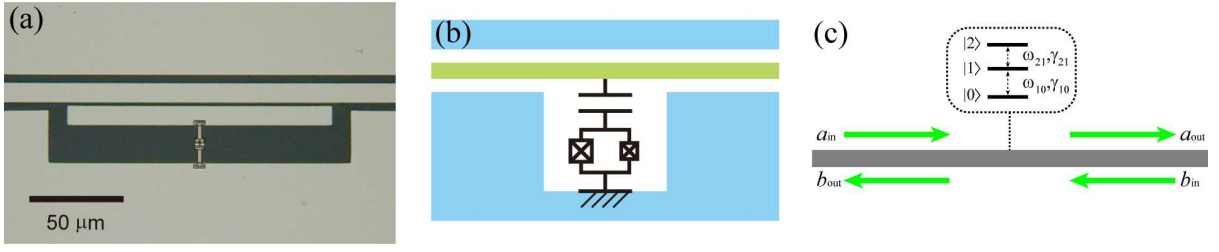}\enc
\caption{
(a)~Optical micrograph of a transmon qubit coupled 
to a microwave transmission line and (b)~its schematic.
(c)~Theoretical model.
A ladder-type three-level system is coupled directly to the transmission line.
The arrows $a$ ($b$) denote the microwave fields propagating rightward (leftward).
}
\label{fig:system}
\end{figure*}

In our setup, a superconducting transmon qubit is coupled directly 
to a one-dimensional microwave-photon field
propagating in a transmission line [Fig.~\ref{fig:system}(a)]. 
The circuits were made using a Josephson-junction fabrication process 
based on NbN/AlN/NbN trilayers described in Ref.~\cite{Nakamura11}. 
A single-island transmon is used in the present work~\cite{tra}. 
The island electrode is capacitively coupled 
to the center line of a 50-$\Omega$ coplanar waveguide 
and is connected to the ground via two parallel Josephson junctions. 
The junctions have different Josephson energies $E_{\rm J1}$ and $E_{\rm J2}$. 
We apply an external magnetic flux of $\Phi_0/2$ ($\Phi_0 = h/2e $) 
through the loop formed by the two junctions, 
so that the effective Josephson energy becomes 
$E_{\rm J}^{\rm eff} = |E_{\rm J1} - E_{\rm J2}|$ 
and the qubit is insensitive to magnetic flux noise to first order. 
From the transition energies obtained in the measurements below, 
the effective Josephson energy and charging energy of 
the transmon are estimated to be $E_{\rm J}^{\rm eff} = 19.05$~GHz 
and $E_{\rm C} = 0.4206$~GHz~\cite{tra}.
When necessary, the qubit frequency can be detuned  
by changing the magnetic flux bias.

Theoretically, our system is described as follows.
We consider the lowest three states of the qubit 
and denote them by $|j\ra$ ($j=0,1,2$), where $\om_0<\om_1<\om_2$. 
The qubit is driven by a continuous field
with amplitude $E$ and frequency $\om_{\rm d}$. 
Setting $\hbar=v=1$, where $v$ is the microwave velocity 
in the transmission line, and working
in a frame rotating at the drive frequency $\om_{\rm d}$, 
the Hamiltonian of the system is 
\bea
\cH &=& \cH_{\rm q}+\cH_{\rm f}+\cH_{\rm q-f},
\label{eq:H}
\\
\cH_{\rm q} &=& (\om_{10}-\om_{\rm d})\s_{11}+(\om_{20}-2\om_{\rm d})\s_{22}
+E(\s_{\rm t}+\s_{\rm t}^{\dagger}),
\label{eq:Hq}
\\
\cH_{\rm f} &=& \int dk \ (ka_k^{\dag}a_k + kb_k^{\dag}b_k),
\\
\cH_{\rm q-f} &=& \frac{1}{\sqrt{2\pi}}
\int dk \left[\s_{\rm t}^{\dagger}(a_k+b_k)+{\rm H.c.}\right],
\eea
where $\s_{ij}=|i\ra\la j|$,
$\s_{\rm t}=\sqrt{\g_{10}/2}\,\s_{01}+\sqrt{\g_{21}/2}\,\s_{12}$ 
is the transition dipole operator of qubit,
and $a_k$ ($b_k$) is the photon annihilation operator 
propagating rightward (leftward) with wave number $k$.
The transition frequency between $|i\ra$ and $|j\ra$
is $\omega_{ij}=\omega_i-\omega_j$,
and $\g_{ij}$ is the radiative decay rate from $|i\ra$ to $|j\ra$.
Note that $\g_{20}=0$ results from a parity selection rule. 
By fitting the experimental results below,
these parameters are estimated as follows (see Appendix~\ref{sec:app1}):
$\om_{10}/2\pi=7.558$~GHz, $\om_{21}/2\pi=7.070$~GHz,
and $\g_{10}/2\pi=40$~MHz.
Based on the transition matrix elements for transmons~\cite{tra}, 
we also assume $\g_{21}={2}\g_{10}$.

\begin{figure*}
\bec\includegraphics[width=170mm]{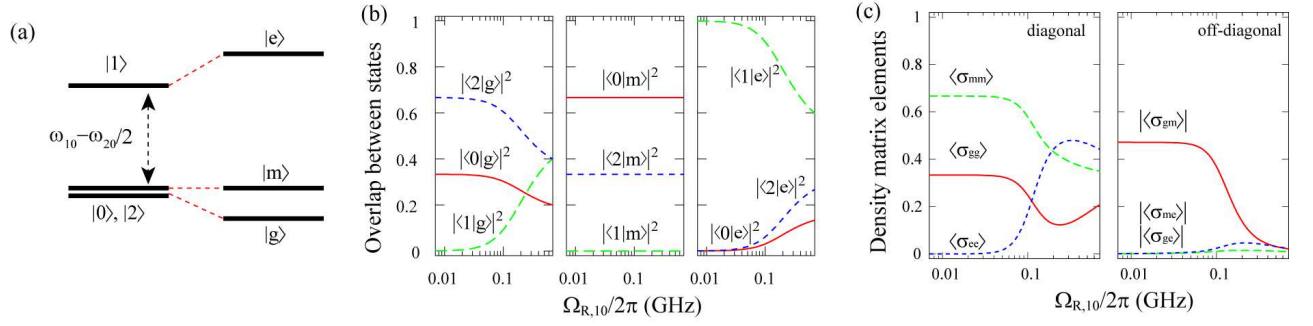}\enc
\caption{
(a)~Level structure in the rotating frame of the bare states (left)
and the dressed states (right).
The drive frequency is set to $\om_{\rm d}=\om_{20}/2$.
(b)~Overlap between the bare and dressed states
as a function of the Rabi frequency. 
(c)~Stationary density matrix elements in the dressed-state basis: 
diagonal elements (left) and off-diagonal elements (right).
}
\label{fig:ls}
\end{figure*}

Due to the weak anharmonicity of transmon qubits,
$\omega_{10}$ is close to $\omega_{21}$. 
Consequently, the three bare states are also energetically close in the rotating frame.
We denote the dressed states, i.e., the (quasi-)eigenstates of $\cH_{\rm q}$,
by $|\mu\ra$ ($\mu=g,m,e$), where $\om_g<\om_m<\om_e$.
We hereafter focus on the case of $\om_{\rm d}=\om_{20}/2$,
where $|0\ra$ and $|2\ra$ are degenerate in the rotating frame [Fig.~\ref{fig:ls}(a)].
The Rabi frequencies for the $|0\ra\leftrightarrow|1\ra$
and $|1\ra\leftrightarrow|2\ra$ transitions are respectively defined by
$\Om_{\rm R,10}=\sqrt{\g_{10}}E$ and $\Om_{\rm R,21}=\sqrt{\g_{21}}E$,
and we use $\Om_{\rm R,10}$ as a measure of the driving field intensity.
Figure~\ref{fig:ls}(b) plots the overlap between the bare and dressed states
as a function of $\Om_{\rm R,10}$.
We observe that $|m\ra$ is composed only of two bare states 
and is insensitive to the drive intensity
($|m\ra=\sin\theta|0\ra-\cos\theta|2\ra$, 
where $\theta=\arctan\sqrt{\g_{21}/\g_{10}}$).
This is specific to the case of $\om_{\rm d}=\om_{20}/2$,
where $|0\ra$ and $|2\ra$ are degenerate.
In general, the dressed states involve the three bare states 
when subjected to a strong driving field, 
where the Rabi frequencies overwhelm the bare-state energy differences
in the rotating frame.
Generation of three-state dressed states requires two fields in general~\cite{EIT4}. 
Here, three-state dressed states can be realized by a single driving field
due to the weak anharmonicity.

From Eq.~(\ref{eq:H}), the Heisenberg equation 
for the qubit is given by 
\bea
\frac{d}{dt}\s_{\mu\nu} &=& i\om_{\mu\nu}\s_{\mu\nu} -\xi_{\mu\nu}
-i\zeta_{\mu\nu} \{ a_{\rm in}(t)+ b_{\rm in}(t) \}
\nonumber \\
& &
+i\{ a^{\dag}_{\rm in}(t)+ b^{\dag}_{\rm in}(t) \}\zeta_{\nu\mu}^{\dag}, 
\label{eq:dsig_dt}
\eea
where 
$\xi_{\mu\nu} = \s_{\mu\nu}\s_{\rm t}^{\dag}\s_{\rm t} + 
\s_{\rm t}^{\dag}\s_{\rm t}\s_{\mu\nu} - 2\s_{\rm t}^{\dag}\s_{\mu\nu}\s_{\rm t}$,
$\zeta_{\mu\nu} = [\s_{\mu\nu},\s_{\rm t}^{\dag}]$,
and $a_{\rm in}$ and $b_{\rm in}$ are the input field operators.
When no probe field is applied, $\la a_{\rm in}\ra=\la b_{\rm in}\ra=0$. 
The stationary density matrix elements $\la\s_{\mu\nu}\ra_{\rm s}$
of the driven qubit are determined by 
\beq
i\om_{\mu\nu}\la\s_{\mu\nu}\ra_{\rm s} 
- \sum_{\mu',\nu'} \xi_{\mu\nu,\mu'\nu'} \la\s_{\mu'\nu'}\ra_{\rm s}=0,
\label{eq:sig_s}
\eeq
where $\xi_{\mu\nu,\mu'\nu'}=\la\mu'|\xi_{\mu\nu}|\nu'\ra$.
These simultaneous equations, 
together with the sum rule $\sum_{\mu}\la\s_{\mu\mu}\ra_{\rm s}=1$,
determine $\la\s_{\mu\nu}\ra_{\rm s}$.
Figure~\ref{fig:ls}(c) plots the density matrix elements
as functions of the Rabi frequency.
When the driving field is weak,
the qubit is in the bare ground state $|0\ra$,
which is a superposition of $|g\ra$ and $|m\ra$.
The qubit is in a pure state in this regime.
As the drive intensity increases,
the qubit becomes entangled with 
the incoherently scattered photons,
and its density matrix becomes progressively more mixed.
As a result, under strong driving, 
the off-diagonal elements become negligibly small
in comparison with the diagonal ones.
In the strong driving limit $E \to \infty$,
the qubit approaches a maximally mixed state, 
where the three diagonal elements each take the value 1/3,
whereas the off-diagonal elements vanish.

\begin{figure*}
\bec\includegraphics[width=170mm]{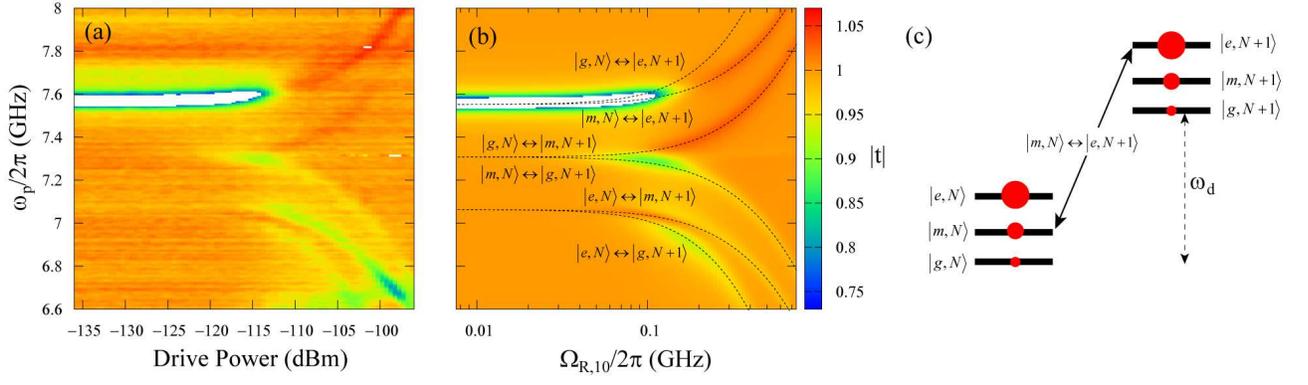}\enc
\caption{
(a)~Experimental and (b)~theoretical plots of transmission $|t|$
as a function of the drive intensity and the probe frequency.
The drive frequency $\om_{\rm d}/2\pi$ is set at 7.314~GHz [$=(\omega_{20}/2)/2\pi$].
In the white region at $\om_{\rm p}/2\pi=7.558$~GHz,  
$|t|$ is less than $0.5$ due to strong reflection. 
In (b), the drive power is expressed in $\Omega_{\rm R,10}$. 
The transition frequencies between the dressed states are superposed, and
they form six Rabi sidebands which appear symmetrically 
with respect to the drive frequency, $\om_{\rm d}/2\pi=7.314$~GHz. 
(c)~Level structure of the dressed states.
The $|m,N\ra \leftrightarrow |e,N+1\ra$ transition is indicated as an example.
Population of the dressed states
[strong drive region of Fig.~\ref{fig:ls}(c)]
is represented by the size of the circles.
}
\label{fig:3d}
\end{figure*}

In order to observe the optical properties of such three-state dressed states,
in addition to the drive microwave, 
we applied a weak continuous probe field 
(amplitude $F$ and frequency $\om_{\rm p}/2\pi$)
from one side of the waveguide
and measured the transmission coefficient $t$.
Experimentally, $t$ is normalized by the transmission coefficient 
obtained at the same frequency 
when the qubit is far-detuned from the probe frequency. 
The probe power at the qubit is fixed at $-$133~dBm, 
which is in the linear regime, $F^2 \ll \gamma_{10}$.
In Fig.~\ref{fig:3d}(a), $|t|$ is plotted as 
a function of the drive power and the probe frequency.
The drive power $P$ is related to the field amplitude by 
$P=\hbar\om_{\rm d}E^2$.
In theory, application of a probe field
from one side is realized by putting 
$\la a_{\rm in}\ra=Fe^{i(\om_{\rm d}-\om_{\rm p})t}$ and $\la b_{\rm in}\ra=0$. 
We investigated the linear response of the qubit since the probe is relatively weak. 
We divide $\la\s_{\mu\nu}(t)\ra$ into the stationary and linear-response components, as 
$\la\s_{\mu\nu}\ra_{\rm s}+\la\s_{\mu\nu}\ra_{\rm L}e^{i(\om_{\rm d}-\om_{\rm p})t}$.
From Eqs.~(\ref{eq:dsig_dt}) and (\ref{eq:sig_s}),
the simultaneous equations used to determine $\la\s_{\mu\nu}\ra_{\rm L}$ are given by
\bea
i(\om_{\mu\nu}+\om_{\rm p}-\om_{\rm d})\la\s_{\mu\nu}\ra_{\rm L} 
- \sum_{\mu',\nu'}\xi_{\mu\nu,\mu'\nu'} \la\s_{\mu'\nu'}\ra_{\rm L}=
\nonumber 
\\ 
iF\times \sum_{\mu',\nu'}\zeta_{\mu\nu,\mu'\nu'} \la\s_{\mu'\nu'}\ra_{\rm s},
\eea
where $\zeta_{\mu\nu,\mu'\nu'}=\la\mu'|\zeta_{\mu\nu}|\nu'\ra$.
The input and output field operators are connected by
$a_{\rm out}(t)=a_{\rm in}(t)-i\s_{\rm t}(t)$.
Therefore, the amplitude of the transmitted wave is given by
\beq
\la a_{\rm out}(t)\ra =
\left( F -i\sum_{\mu,\nu}\s_{{\rm t},\mu\nu}\la\s_{\mu\nu}\ra_{\rm L}\right)
e^{i(\om_{\rm d}-\om_{\rm p})t},
\eeq
where $\s_{{\rm t},\mu\nu}=\la\mu|\s_{\rm t}|\nu\ra$.
The transmission coefficient is defined as 
$t=\la a_{\rm out} \ra/\la a_{\rm in}\ra$, and 
Fig.~\ref{fig:3d}(b) plots $|t|$ thus calculated. 
Discrepancy between the experiment and theory is found at
$\om_{\rm p}/2\pi \lesssim 6.8$~GHz. 
This is due to the third bare state $|3\ra$,
the transition energy of which is supposed to lie around here.
Except for this point, 	
we confirm good agreement between the experiment and theory
despite the simple model 
neglecting dephasing and non-radiative decay of the qubit.
This indicates the realization of coherent and almost perfect radiative coupling
between the qubit and microwave fields in our setup.

In the weak-drive region ($\Om_{\rm R,10} \lesssim |\om_{10}-\om_{20}/2|$),
the bare-state picture is valid.
Since the probe field is weak, we may regard the qubit 
as a two-level system having two states, $|0\ra$ and $|1\ra$.
The qubit then reflects with high efficiency the probe waves 
at $\om_{10}/2\pi$=7.558~GHz 
due to the one-dimensionality of the field~\cite{Ast}.
In contrast, in the strong-driving regime
($\Om_{\rm R,10} \gtrsim |\om_{10}-\om_{20}/2|$),
the Rabi splittings become observable 
and transitions occur between the dressed states.
In the dressed-state picture,
as illustrated in Fig.~\ref{fig:3d}(c),
optical transitions occur between $|\mu,N\ra$ (lower levels) 
and $|\nu,N+1\ra$ (upper levels),
where $\mu,\nu=g,m,e$ and $N$ denotes the drive photon number.
Therefore, the three-state dressed states
yield six Rabi sidebands at $\om_{\rm d}+\om_{\nu}-\om_{\mu}$ ($\mu\neq\nu$).
Remarkably, we observed all of six sidebands in our experiment
through amplification or attenuation of the probe field [Fig.~\ref{fig:3d}(a)].
In theory, when the probe is tuned to 
the $|\mu,N\ra \leftrightarrow |\nu,N+1\ra$ transition $(\mu\neq\nu)$,
at a frequency $\om_{\rm p}=\om_{\rm d}+\om_{\nu}-\om_{\mu}$,
the transmission coefficient $t$ is reduced to the following form:
\beq
t=1+\frac{|\la\mu|\s_{\rm t}|\nu\ra|^2}{\xi_{\mu\nu,\mu\nu}}
(\la\s_{\nu\nu}\ra_{\rm s}-\la\s_{\mu\mu}\ra_{\rm s}),
\eeq
where $\xi_{\mu\nu,\mu\nu}$ is a positive quantity
of the order of radiative decay rates. 
Therefore, the probe field is amplified (attenuated)
when the population is inverted (not inverted) 
in the dressed-state basis.
We can check in Fig.~\ref{fig:ls}(c) that 
$\la\s_{ee}\ra > \la\s_{mm}\ra > \la\s_{gg}\ra$ under a strong driving,
and this is represented in Fig.~\ref{fig:3d}(c)
by the size of the circles.
For the $|m,N\ra \leftrightarrow |e,N+1\ra$ transition for example,
the upper level is more populated than the lower level. Namely,
the population is inverted 
and therefore the probe field is amplified in this case.
In contrast, for its counterpart transition 
of $|e,N\ra \leftrightarrow |m,N+1\ra$, 
the population is not inverted and the probe field is attenuated.
As a result, in Fig.~\ref{fig:3d}(a) and Fig.~\ref{fig:3d}(b),
amplification and attenuation appear symmetrically 
with respect to the drive frequency $\om_{\rm d}/2\pi$=7.314~GHz.
In principle, transitions between the same dressed states,
$|\mu,N\ra \leftrightarrow |\mu,N+1\ra$, may also take place,  
since the transition dipole does not necessarily vanish.
However, in this case [see Fig.~\ref{fig:3d}(c)], 
the upper and lower levels are equally populated 
and the qubit correspondingly becomes transparent. 
We observe in Fig.~\ref{fig:3d}(a) and (b)
that $|t|=1$ at $\om_{\rm p}=\om_{\rm d}$ under strong driving conditions.
In Appendix~\ref{sec:om10}, we present the results for another drive frequency
($\om_{\rm d}=\om_{10}$) to further support
the validity of the above arguments.

To summarize, we have investigated the microwave response 
of a driven transmon qubit coupled directly to a transmission line.
Owing to the weak anharmonicity of a transmon qubit,
we generate dressed states using a single driving field
that comprise three bare states.
We confirmed formation of these three-state dressed states 
by observing the amplification and attenuation 
of the probe field at six Rabi sidebands. 
The experimental results are reproduced with good precision
by a theoretical model
considering only the radiative coupling between the qubit and the transmission line.
This indicates realization of 
a clean and lossless one-dimensional quantum optical system
that is suitable for constructing scalable quantum networks.

The authors are grateful to R. Yamazaki, 
W. D. Oliver and O. Astafiev for discussion.
This work was partly supported by 
the Funding Program for World-Leading Innovative R\&D 
on Science and Technology (FIRST), 
Project for Developing Innovation Systems of MEXT, 
MEXT KAKENHI (Grant Nos. 21102002, 25400417),
SCOPE (111507004),
and National Institute of Information and Communications
Technology (NICT).

\appendix
\section{estimation of $\bm{\gam_{10}}$}
\label{sec:app1}
Here we investigate the power dependence of the transmission coefficient
and estimate the qubit parameters. 
For simplicity, we investigate transmission of the drive wave alone,
without applying the probe wave.
We assume that the drive is tuned to the lowest transition 
of the qubit ($\om_{\rm d}\simeq\om_{10}$).
Then, in the weak-power limit, 
we can treat the transmon qubit as a two-level system. 
Furthermore, for better fitting of the experimental data, 
we include nonradiative decay and pure dephasing of the qubit here.
We add the following terms to the Hamiltonian [Eq.~(1) of the main text]: 
\bea
\cH_{\rm n}+\cH_{\rm q-n}
&=&
\int dk \left[k c_k^{\dag}c_k+
\sqrt{\gam'_{10}/2\pi}\,(\s_{10}c_k+c_k^{\dag}\s_{01})
\right],
\\
\cH_{\rm p}+\cH_{\rm q-p}
&=&
\int dk \left[k d_k^{\dag}d_k+
\sqrt{\gam_{\rm p}/\pi}\,\s_{11}(d_k+d_k^{\dag})
\right],
\eea
where the environmental degrees of freedom are modeled 
by boson fields ($c_k$ and $d_k$)
and $\gam'_{10}$ and $\gam_{\rm p}$ respectively denote
the rates for nonradiative decay and pure dephasing.
The equations of motion for $\la\s_{01}\ra$ and $\la\s_{11}\ra$ are given by
\bea
\frac{d}{dt}\la\s_{01}\ra
&=&
-(\Gam_2-i\delta\om)\la\s_{01}\ra
-i\frac{\Om_{\rm R}}{2}(1-2\la\s_{11}\ra),
\\
\frac{d}{dt}\la\s_{11}\ra
&=&
-\Gam_1^{\rm tot}\la\s_{11}\ra
+i\frac{\Om_{\rm R}}{2}(\la\s_{01}\ra-\la\s_{01}\ra^*),
\eea
where $\Gam_1=\gam_{10}$, 
$\Gam_1^{\rm tot}=\gam_{10}+\gam'_{10}$,
$\Gam_2=\Gam_1^{\rm tot}/2+\gam_{\rm p}$, and
$\delta\om=\om_{\rm d}-\om_{10}$.
The Rabi frequency $\Om_{\rm R,10}$ is denoted here by $\Om_{\rm R}$.
Note that the expectation values of the input field operators ($\la a_{\rm in}\ra$, etc.)
disappear since no probe wave is applied here.
The transmission coefficient is given by
$t=1-i\sqrt{\gam_{10}}\la\s_{01}\ra_{\rm s}/E$,
where $\la\s_{01}\ra_{\rm s}$ is 
the stationary solution of $\la\s_{01}\ra$. 
Therefore, we have
\beq
t=1-\frac{\Gam_1}{2\Gam_2}\frac{1+i\delta\om/\Gam_2}
{1+(\delta\om/\Gam_2)^2+\Om_{\rm R}^2/\Gam_1^{\rm tot}\Gam_2}.
\label{eq:t}
\eeq

\begin{figure}
\bec\includegraphics[width=110mm]{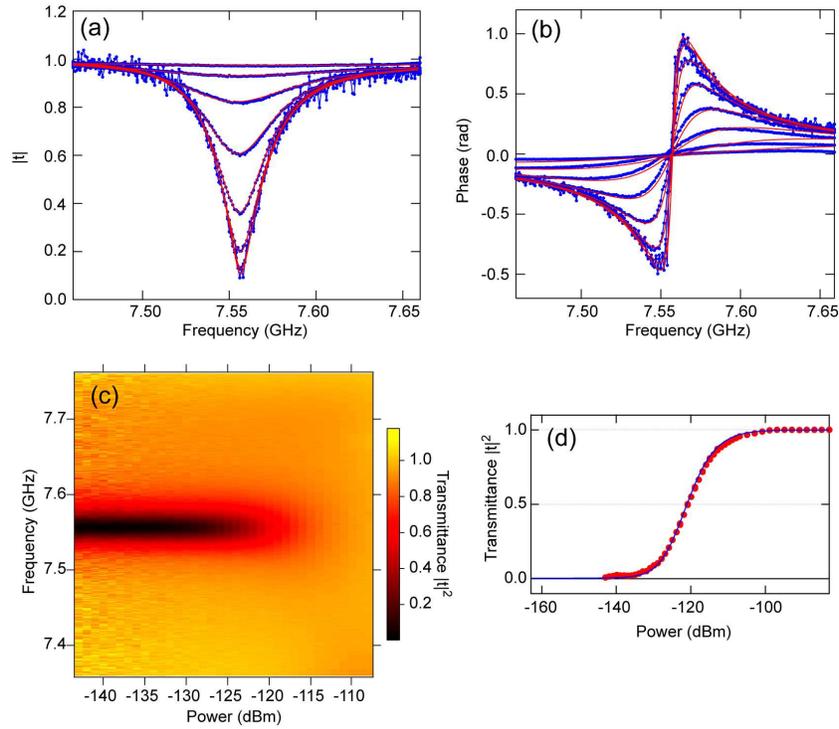}\enc
\caption{
Characterization of resonance fluorescence of the transmon.
(a)~Amplitude and (b)~phase of the transmission coefficient $t$
as functions of the drive frequency.
The drive power is $-108$~dBm for the top curve 
and decreasing with 5-dB step
from top to bottom (blue: experiment, red: theory).
There are eight curves plotted, while the two lowest curves are mostly overlapping. 
(c)~ Intensity plot of the transmittance 
as a function of the drive power and frequency. 
(d)~Transmittance at the resonant frequency 
as a function of the drive power. 
Circles are experimental data points and the line is the fit.
}
\label{fig:A1}
\end{figure}

In Fig.~\ref{fig:A1},
the amplitude and the phase of the measured transmission coefficient
are plotted as functions of the detuning
for eight different drive powers.
Furthermore, the transmittance as a function 
of the drive power and frequency is plotted.
We observe that the transmission wave vanishes 
nearly completely for a resonant and weak drive.
This is due to destructive interference 
between the input wave and radiation from the qubit~\cite{Ast},
and indicates excellent one-dimensionality of this setup.
As the drive power increases,
transmission increases due to saturation of the qubit.
By fitting these experimental data by Eq.~(\ref{eq:t}),
we estimate that 
$\gam_{10}/2\pi=40$~MHz, $\gam'_{10}/2\pi=0.5$~MHz, 
and $\gam_{\rm p}/2\pi=1$~MHz.
We also calibrated the drive power delivered to the qubit 
against the Rabi frequency $\Omega_{\rm R}$ by the fitting of Fig.~\ref{fig:A1}(d).

\section{Results for $\bm{\om_{\rm d}=\om_{10}}$}
\label{sec:om10}
\begin{figure}
\bec\includegraphics[width=140mm]{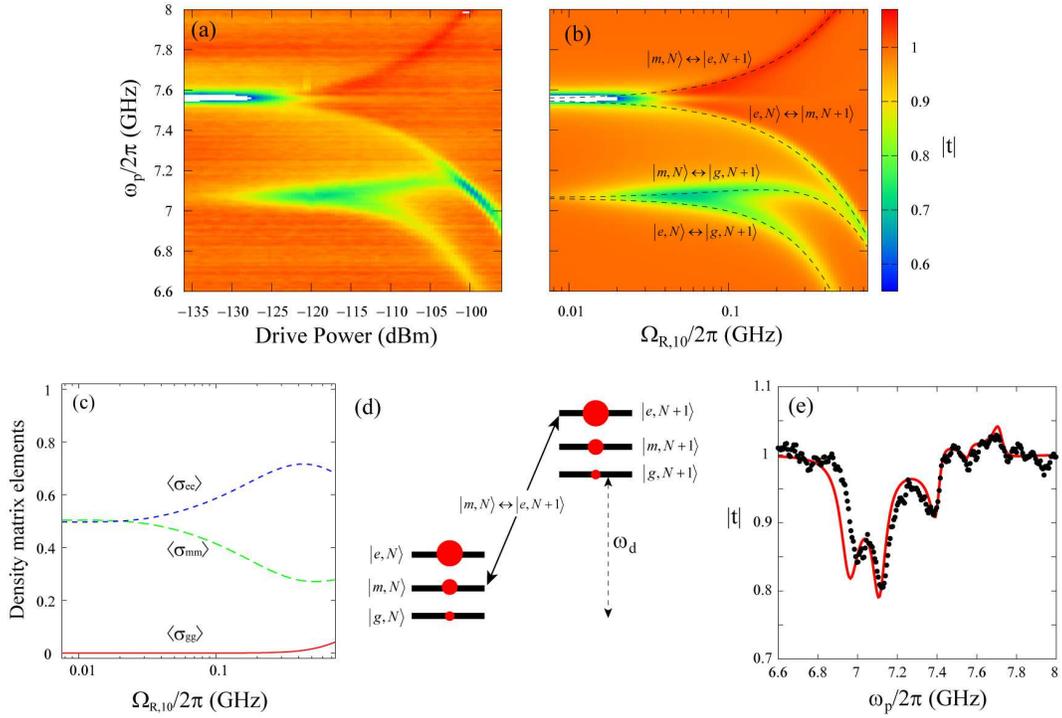}\enc
\caption{
Spectroscopy of the dressed states for the resonant drive 
at $\omega_{\rm d} = \omega_{10}$. 
(a)~Experimental and 
(b)~theoretical plots of $|t|$.
The transition frequencies between the dressed states are superposed.
(c)~Population of the dressed states as functions of the drive power.
(d)~Level structure of the dressed states.
Circles represent the population of the dressed states
[strong drive region of (c)].
(e)~Cross section of (a) and (b) under a fixed drive power
($-$110~dBm, where $\Om_{\rm R,10}/2\pi=0.113$~GHz).
Dots (line) represent the experimental (theoretical) results.
}
\label{fig:A2}
\end{figure}

In the main text, the drive frequency is fixed at $\om_{\rm d}=\om_{20}/2$. 
Here we present the results for 
$\om_{\rm d}=\om_{10}$ in Fig.~\ref{fig:A2} for reference.
This figure shows
(a)~experimental and (b)~theoretical plots of $|t|$,
(c)~population and (d)~energy diagram of the dressed states,
and (e)~cross section of (a) and (b) under a fixed drive power.
We observe that the experimental results are reproduced well by the theory.

In the weak-drive regime,
optical transitions occur between the bare states, $|0\ra$ and $|1\ra$.
We observe strong attenuation of the probe wave at $\om_{\rm p}=\om_{10}$,
which is due to the nearly perfect reflection of the input wave.
This signal is observable regardless of the drive frequency [see Fig.~3(a)]. 
In contrast, in the strong-drive regime,
transitions between the dressed states become observable.
When $\om_{\rm d}=\om_{10}$, 
the drive wave mixes $|0\ra$ and $|1\ra$ 
to constitute the dressed states $|m\ra$ and $|e\ra$
whereas $|g\ra \simeq |2\ra$.
The qubit population satisfies
$\la\s_{ee}\ra>\la\s_{mm}\ra>\la\s_{gg}\ra$, 
as shown in Fig.~\ref{fig:A2}(c).
As a result, the probe is amplified when tuned to the 
$|m,N\ra \leftrightarrow |e,N+1\ra$ transition,
and is attenuated when tuned to the 
$|e,N\ra \leftrightarrow |m,N+1\ra$,
$|m,N\ra \leftrightarrow |g,N+1\ra$
and $|e,N\ra \leftrightarrow |g,N+1\ra$ transitions.

\end{document}